\documentclass[useAMS,usenatbib]{mn2e}

\title[Molecular cloud formation in a two phase medium]
{Spiral shocks and the formation of molecular clouds in a two phase medium}
\author[C. L. Dobbs, I. A. Bonnell]
{C. L. Dobbs$^{1,2}$\thanks{E-mail:
dobbs@astro.ex.ac.uk} \& I. A. Bonnell$^1$ \\
$^1$ SUPA, 
School of Physics and Astronomy, University of St Andrews, 
North Haugh, St Andrews, Fife, KY16 9SS \\
$^2$ School of Physics, University of Exeter, 
Stocker Road, Exeter, EX4 4QL}
\usepackage{amssymb}
\usepackage{amsmath}
\usepackage{psfig}

\begin{document}
\date{\today}

\pagerange{\pageref{firstpage}--\pageref{lastpage}} \pubyear{0000}

\maketitle

\label{firstpage}

\begin{abstract}
We extend recent numerical results \citep{DBP2006} on molecular cloud 
formation in spiral
galaxies by including a multi-phase medium. The addition of a hot phase of gas
enhances the structure in the cold gas, and significantly
increases the fraction of molecular hydrogen that is formed when the cold gas
passes through a spiral shock.
The difference in structure is reflected in the mass power spectrum of the
molecular clouds, which is
steeper for the multi-phase calculations. The increase in molecular gas occurs
as the addition of a hot phase leads to higher densities in the cold gas. 
In particular, cold gas is confined in clumps between the
spiral arms and retains a higher molecular fraction. Unlike the single
phase results, 
molecular clouds are present in the inter-arm regions for the multi-phase
medium. However the density of the
inter-arm molecular hydrogen is generally below that which can be reliably
determined from CO measurements. We therefore predict that for a multi-phase medium, 
there will be low density clouds containing cold atomic and molecular hydrogen, 
which are potentially entering the spiral arms. 
\end{abstract}


\section{Introduction}
Numerical simulations have become a powerful tool to explore the physics
of star formation \citep{Ostriker2001,Bate2003,Maclow2004,BPPV2006}.
Recent advances in computational power now make it possible to study the formation of
giant molecular clouds (GMC) and the onset of star formation \citep{Kim2003,Bonnell2006,DBP2006,Glover2006a,
Glover2006b,Vaz2006}. 
There are two general formation mechanisms that have been advanced. the first is that
a cloud becomes self-gravitating, inducing high-densities and the conversion of  HI into molecular gas.
The second possibility is that an external triggering such as a spiral or other turbulent shock 
either gathers together pre-existing molecular gas \citep{Pringle2001}  or induces high densities
in the shock and thus converting HI into H2.

Previous results show that spiral shocks induce the formation of dense
molecular clouds \citep{DBP2006} and simultaneously 
generate a velocity dispersion \citep{Bonnell2006}. 
A prerequisite for the formation of these molecular clouds 
in isothermal simulations is that the gas is cold, of order 100 K or less 
\citep{DBP2006}. For cold gas, the densities in the shocks are sufficiently 
high for H$_2$ formation, and the dynamics of the spiral shocks lead to clumpy
structures in the gas. 
The cold atomic component of the ISM 
contains the most mass of atomic hydrogen in the ISM, but the
filling factor of this phase is very small ($\sim 0.013$, \citealt{Cox2005}). 
The warm ($10^4 $K) atomic phase, which is typically used in
simulations of spiral galaxies \citep{Kim2002,Wada2004,Gittins2004},
contains less mass but fills just over a third of the
volume of the ISM \citep{Cox2005}.   
The simulation described in this paper is the first to examine the dynamics of
a spiral potential with both cold and warm phases. 
The resulting gas distribution is more representative of 
the ISM, with cold gas embedded in smaller clumps within a warm diffuse phase. 

In the present study, we investigate the effects of the hot medium on  
a pre-existing cold
HI as it passes through the spiral shock. The motivation is to  
include the pressure
confinement of the hot gas in the dynamics of a spiral shock. We do not
include any  transition between the hot and cold components but  
rather assume the picture presented by \citep{Pringle2001} that molecular clouds  
are formed from gas that is cold before it enters the shock. We leave a full treatment of the  
thermal evolution of the gas to a later study.
\citet{Glover2006a,Glover2006b} analyze molecular cloud formation 
over much smaller size-scales (20-40 pc), but include a 
more consistent treatment of the ISM and H$_2$ formation.
In their models, molecular clouds form slowly in gravitationally
collapsing uniform gas \citep{Glover2006a} or rapidly from the compression of turbulent
 gas  \citep{Glover2006b}. The gas temperature in these calculations
is initially  $10^4$ K, but  quickly drops to 100 K or less once the density
exceeds 1 cm$^{-3}$. Other models of the ISM have shown the formation of cold 
($<100$ K) HI, as a precursor
to molecular clouds, on local \citep{Vaz1995,Audit2005,Heitsch2005} and
galactic \citep{Wada1999} scales.

We present numerical results to investigate the dynamics of cold gas subject to a spiral shock when 
in the presence of a confining warm phase. 
We compare this 2 phase simulation of  a galactic disc with the 
previous single phase simulations shown in \citet{DBP2006}.
In particular, we describe the structure of molecular clouds and the content of
molecular hydrogen across the galactic disc.
We also provide the first attempts to determine individual molecular 
cloud properties from global simulations.  

\section{Calculations}
We use the 3D smoothed particle hydrodynamics (SPH) code based on the version by
Benz \citep{Benz1990}. The smoothing length is allowed to vary with space and
time, with the constraint that the typical number of neighbours for each particle 
is kept near $N_{neigh} \thicksim50$.  
Artificial viscosity is included with the standard parameters $\alpha=1$
and $\beta=2$ \citep{Monaghan1985,Monaghan1992}. 

\subsection{Flow through galactic potential}
The galactic potential includes spiral components for the disc, dark matter halo
and spiral density pattern. The potential is described in full in
\citet{DBP2006}. The spiral component has 4 spiral arms and is taken
from \citet{Cox2002}.
The amplitude is 1~atom~cm$^{-3}$,
the pattern speed is $2 \times 10^{-8}$~rad~yr$^{-1}$ and 
pitch angle $\alpha =15^o$. 
The pattern speed leads to a co-rotation radius of 11kpc. 
The disc is in equilibrium, as the rotational velocities of the gas balance the
centrifugal force of the potential.
 
Self-gravity magnetic fields,
heating, cooling or feedback from star formation are not included in these
calculations.  We further neglect pressure due to cosmic rays, which contributes
tot the pressure of the diffuse ISM. 
The paper instead focuses on how hydrodynamic forces and the
galactic potential influence the flow.
 
\subsection{Initial conditions and calculation of molecular gas density}
Gas particles are initially distributed as described in
\citet{DBP2006}, occupying a region of radius
5 kpc$<r<$10 kpc. The disc has a scale height $z\leq 100$ pc, and a velocity
dispersion of 2.5\% of the orbital speed for the $z$ component of the velocities
maintains vertical equilibrium.  
We determine the $x$ and $y$ components of the positions and velocities 
from a 2D test particle run. Particles in the test run are 
initially distributed uniformly with
circular velocities. They evolve for a couple of orbits subject to the
galactic potential, to give a spiral density pattern with particles settled into
their perturbed orbits. A velocity dispersion of 2.5\% of the orbital speed is 
also added to the $x$ and $y$ velocities. 

The gas is distributed in 2 phases, a cold component of $T=100$ K, and a warm 
component of $T=10^4$ K. Each phase constitutes half of the gaseous mass of 
the disc, $5\times 10^8$ M$_{\odot}$, so the total mass of the disc is 
$10^9$ M$_{\odot}$.
This corresponds to a nominal surface density of $\sim 4$ M$_{\odot}$ pc$^{_2}$
($9 \times 10^{-4}$ g cm$^{-2}$).
The total HI surface density at the solar radius is 
$\sim5$ M$_{\odot}$ pc$^{_2}$ \citep{Wolfire2003}.
The SPH particles are initially randomly assigned as hot or cold gas, but the 
distribution settles into a diffuse warm and dense cold phase as the simulation 
progresses. The gas is isothermal for both phases.  
The calculation of the smoothing lengths
does not distinguish between cold and hot particles. Using two phases can lead to numerical issues with the smoothing lengths \citep{Thacker2000}, in particular overcooling, although both phases are isothermal in our simulations. Furthermore only a small fraction of cold particles have predominantly hot neighbours e.g. only 2 \% of cold particles have $<20$\% cold neighbours. 

Ideally, a full treatment of the ISM, which includes heating and cooling between 
the two phases should be applied. 
However, the primary aim of this paper is to 
investigate how a hot phase affects the dynamics of cold gas passing through a 
spiral shock. There have been recent theoretical arguments 
that molecular clouds may form from cold HI or molecular gas \citep{Pringle2001}, motivated by observational evidence for shorter cloud 
lifetimes. Observations also  imply that molecular clouds form from molecular gas in 
M51 \citep{Vogel1988}, and thre may be reservoirs of 
low density H$_2$ present in galaxies which is difficult to detect 
\citep{Lequeux1993,Allen2004}. Cold HI is ubiquitous in the Outer 
 Galaxy \citep{Gibson2006} and appears to be flowing into the spiral arms 
 \citep{Gibson2005}.  
      
We calculate the density of molecular hydrogen by the same process as described
in \citet{DBP2006}. We post-process our results to determine the fraction of 
molecular gas, using the equation for the rate of change of H$_2$ density from
\citet{Bergin2004}
\begin{equation}
\frac{dn(H_2)}{dt}=R_{gr}(T)n_p
n(H)-[\zeta_{cr}+\zeta_{diss}(N(H_2),A_V)]n(H_2).
\end{equation} 
The total number density is $n_p$, where $n_p=n(H)+2n(H_2)$, 
$N$ is the column density (of atomic or molecular hydrogen) and $T$ the 
temperature. 
The strength of the UV field is $\zeta_{diss}(0)=4.17 \times 10^{-11}$ s$^{-1}$,
comparable with the UV flux from a B0 star, and the cosmic ray ionization rate is
$\zeta_{cr}=6 \times 10^{-18}$ s$^{-1}$.
There is some uncertainty in the cosmic ray ionization rate (e.g. \citealt{LePetit2004})
but this term is much less than the the other components of Eqn.~1 
and unlikely to affect our results.
$R_{gr}$ is the formation rate on grains, assuming an efficiency of
$S=0.3$. We do not explicitly change $S$ for the warm gas in
our calculations, but the density of the warm gas is too low for H$_2$ formation
to occur \citep{DBP2006} and H$_2$ is dissociated at high temperatures. 
Ideally, a more realistic treatment of the ISM
including heating and cooling should be included for a more accurate
calculation of H$_2$ formation. However, as shown in \citet{DBP2006}, this
method primarily extracts the densest gas. 

Our main simulation was performed with 6 million particles, 3 million in each of the hot and cold phases.
This compares to our previous single-phase simulation of 4 million cold particles \citep{DBP2006}.
In addition to these high resolution runs, we ran lower resolution runs (of 1 million and 250,000 particles)
to investigate the effects of the numerical resolution. The lower resolution runs generally yield lower
maximum densities throughout and lower H$_2$ fractions. 
\begin{figure}
\centering
\begin{tabular}{c}
\psfig{file=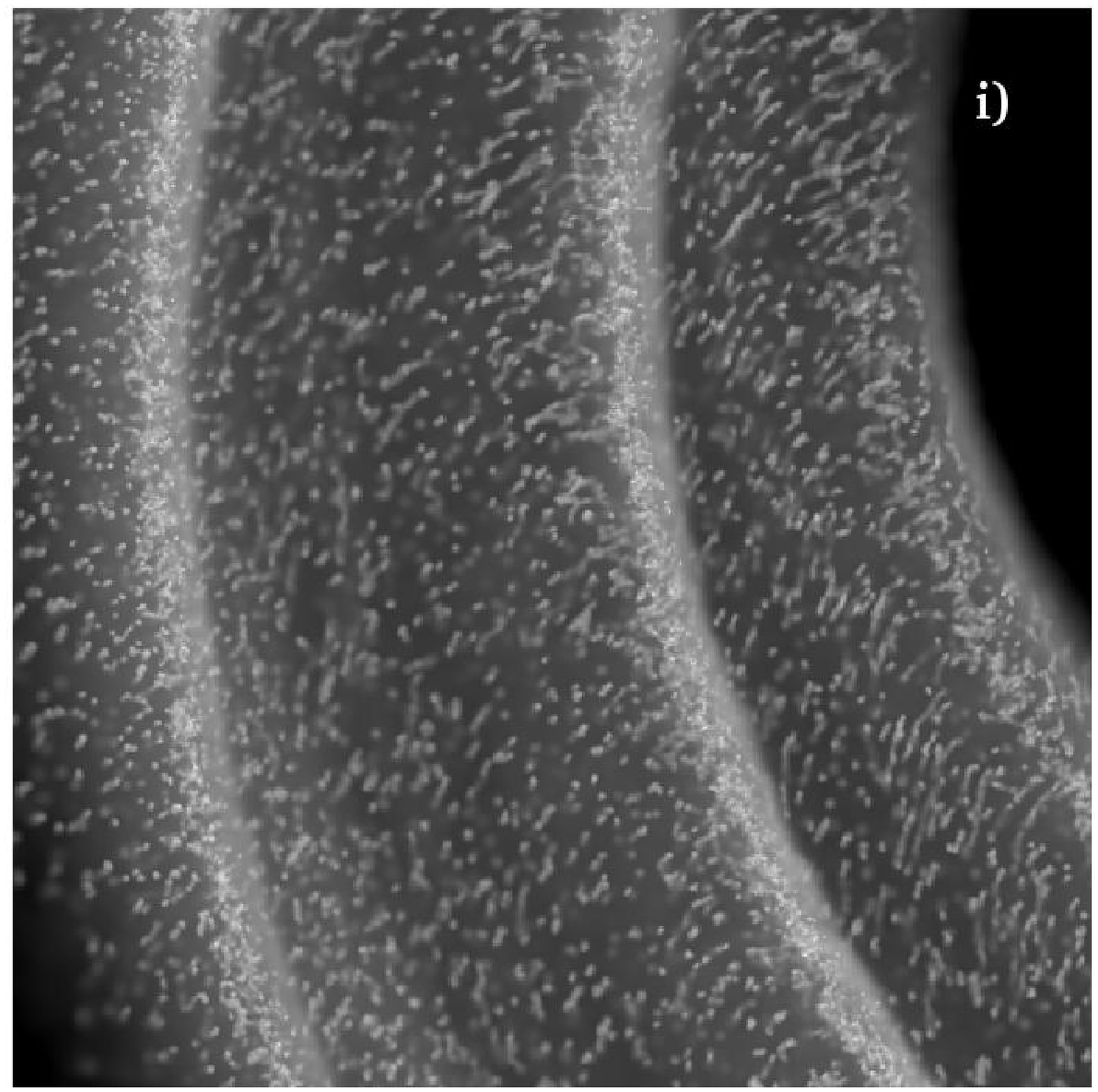,width=2.2in} \\
\psfig{file=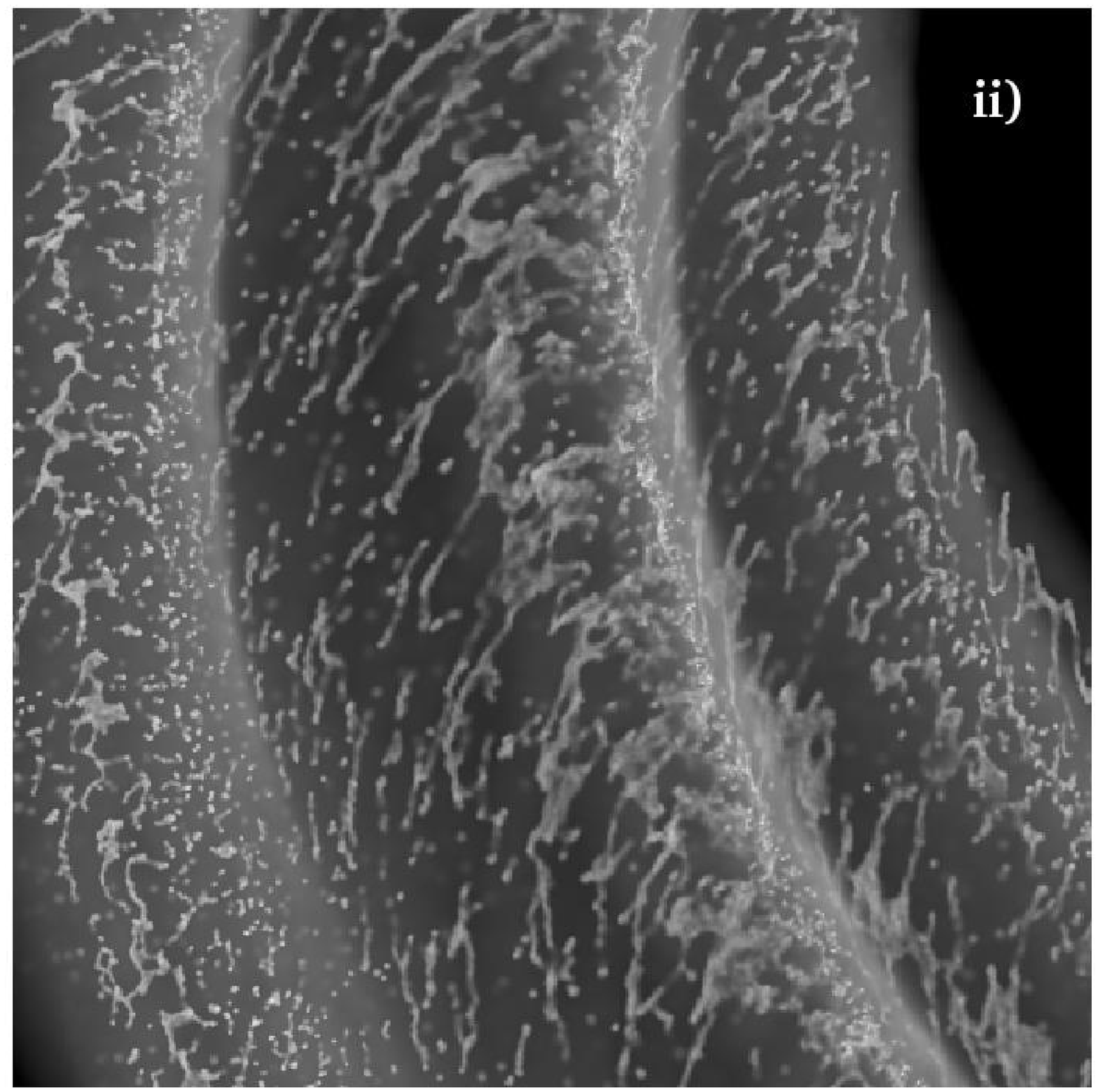,width=2.2in} \\
\psfig{file=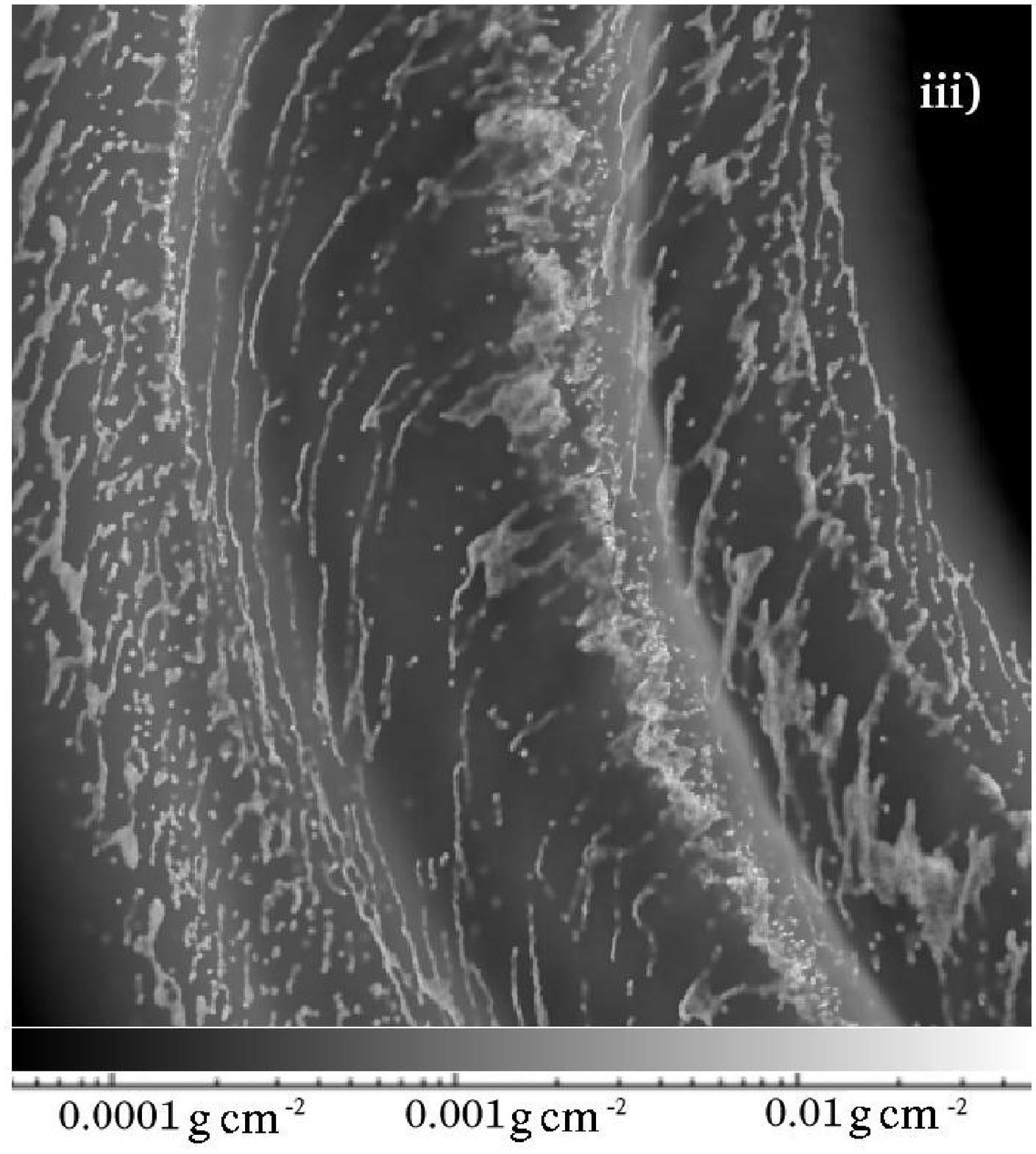,width=2.2in}
\end{tabular}
\caption{Column density plots (g~cm$^{-2}$) showing a 5 kpc by 5 kpc section of 
the disc (with Cartesian coordinates -9 kpc $< x <$ -4 kpc and -5 kpc $< y <$
0 kpc. The xy-coordinate grid is centred on the midpoint of the disc and remains
fixed with time. 
Gas is flowing clockwise across the disc, i.e. from bottom to top.
The time corresponding to each plot is i)~60~Myr, ii)~140~Myr and iii)~220~Myr.
The plot shows the total column density, including both the 100 and 10$^4$ K components.}
\end{figure}

\section{Results}
We first discuss the overall structure of the disc and the overall distribution
of molecular hydrogen. In Section~3.5, we use a clump-finding
algorithm to determine the properties of individual molecular clouds. We compare
results from the multiphase calculation and the single phase (50~K) simulation
described in \citet{DBP2006}. In Sections~3.3 and 3.6.1 we compare the inter-arm and arm
molecular gas densities for the multi-phase simulation with observational results,
and comment on the ratio of inter-arm to arm molecular clouds.

\subsection{Structure of the disc}
The column density section of the disc of the multi-phase simulation
is shown in Fig.~1 at different times. 
From Fig.~1, we see the growth of small scale structures into self
consistent, large scale features over time.
The figure includes both the 100 and $10^4$ K gas. The 100 K gas is
most clearly visible in small clumps whilst the $10^4$ K gas provides a diffuse
background and the smooth component of the spiral arms.
The 2 components are shown explicitly in Fig.~2, which plots the
particles representing the hot and cold gas separately for a section of the disc
after 100~Myr. Both figures show that the cold gas is
confined by the pressure from the hot component into small dense clumps. The hot
gas on the other hand is much more uniform and smooth. 
As previously described \citep{DBP2006}, the structure of the gas in the spiral
arms is dependent on the gas dynamics of the spiral shock. For cold gas,
the spiral shock increases clumpy structure in the gas. For hot gas the shock 
is much weaker, so has much less effect on the dynamics, and the gas pressure
will smooth out any structure in the gas \citep{Dobbs2006}.
Two further differences are apparent for the spiral shock for each component.
As can be seen in Fig.~2, the shock
for the cold gas is offset from the shock corresponding to the hot gas.
However the shock from the hot component is evidently affecting the cold gas
clumps, as can be seen from the elongation of the cold gas clumps as they enter
the spiral arm (Fig.~2).
The shock for the hot gas is also much wider than for the cold gas, since the degree 
of compression of the hot gas is much less, whilst the cold gas forms a very narrow 
spiral arm. 

Fig.~3 shows the distribution of mass with density for the multi and single 
phase simulations after 100 Myr. 
The hot component contributes a greater mass at low densities for the 
multi-phase simulation. The greater mass
at high densities, compared to the single-phase disc is due to the confinement
of the cold gas to higher densities by the hot component.  

Whilst the distribution of hot gas changes little with time, the structure of
the cold gas varies. 
Until approximately 60 Myr (Fig.~1i), 
the cold gas is fairly regularly distributed in
small clumps across the section of disc, with an increased concentration in the
spiral arms.
Unlike the single-phase simulations, where the gas is clumpy due to
the particle nature of the code, the warm phase here confines the cold gas into
dense clumps. 
For the single-phase calculations, the clumpiness in the spiral arms is 
amplified
with time by the dynamics of the shock \citep{DBP2006}.
In these multi-phase calculations however, 
the spiral shock acts rather to organise these clumps into
more coherent structures. These are perhaps more easily visible as the larger
features shearing away from the spiral arms in Fig.~1ii and iii.
By 220 Myr (Fig.~1iii) the 100 K spiral arm clumps have been sheared into thin 
filaments between the arms, whilst new structures are breaking away from the 
spiral arm.
\begin{figure}
\centering
\psfig{file=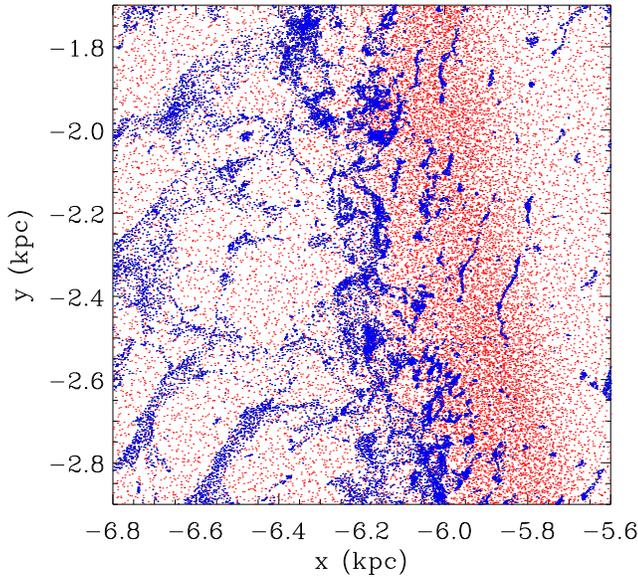,height=3.2in} 
\caption{The particle positions are shown for a section of the disc, 
assuming a Cartesian grid centred on the midpoint of the disc. The 2 phases are
plotted, 100 K (blue) and $10^4$ K (red) after 100 Myr.}
\end{figure}

\begin{figure}
\centering
\psfig{file=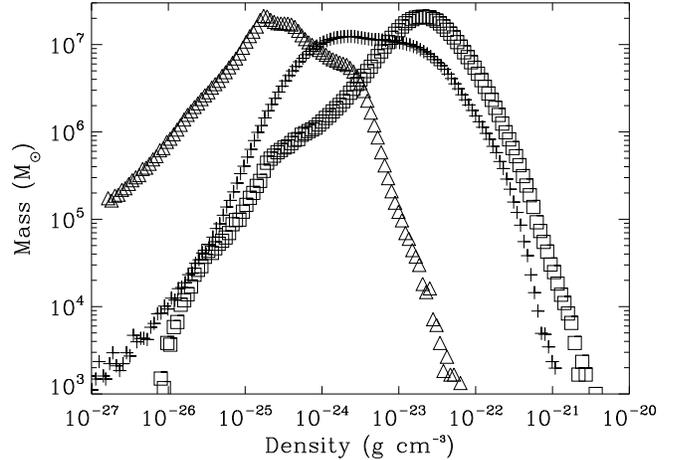,height=2.7in} 
\caption{The distribution of mass as a function of gas density is shown after
100 Myr for the single (crosses) and multi-phase (100 K, squares and $10^4$ K
gas, triangles) simulations. The nominal density for 
molecular clouds is n(H$_2$)=100~cm$^{-3}$ or $3 \times
10^{-22}$ g cm$^{-3}$ \citep{Williams2000}.}
\end{figure}

\subsection{Molecular gas density}
We show column density plots for the simulation after 100 Myr in Fig.~4.
The column density of molecular hydrogen is highlighted in red. 
Fig.~4 shows more clearly the difference in structure of
the cold gas in the single and multi-phase simulations, in particular by
comparing
Fig.~4 iv) and the equivalent frame
of the single-phase simulation (Fig.~9 iv, \citealt{DBP2006}). The multi-phase 
gas is much more structured, with
many smaller scale clumps visible. There are also small clumps of cold gas
strewn across the inter-arm regions. Gas in the single-phase inter-arm regions is
sheared into spurs and feathering \citep{Dobbs2006}, but individual
clumps and structures are much more defined in the multi-phase medium. The cold gas
assembles into longer, more coherent, very thin spurs in the inter-arm regions 
at later times in the multi-phase calculation (Fig.~1iii). 

The molecular
hydrogen corresponds to 12\% of the total mass, i.e. 1.2 $\times 10^8$
M$_{\odot}$ and 24\% of the cold gas. Most of the clumps of cold gas are 
over-plotted with the molecular hydrogen column density. 
The most striking difference between these results and those of
\citet{DBP2006} (Fig.~9) is the greater amount of molecular
gas. 
The fraction of molecular gas is approximately twice that of the single-phase
simulation after 100 Myr.
Unlike the single-phase calculations, the cold gas is confined by the
pressure from the hot gas and remains
more dense in the inter-arm regions. Consequently there is less photodissociation
of H$_2$, and higher densities of molecular gas are produced between the spiral
arms. As seen from Fig.~4 iii) and iv), the molecular gas does not become 
fully dissociated
between the spiral arms, unlike the single-phase medium.
The mass of molecular gas continues to rise
throughout the multi-phase simulation, unlike the single-phase results, 
which peaked after
about 100 Myr. As the gas is not fully dissociated between the arms, the total
molecular gas density continues to increase after multiple spiral arm passages.
By the end of the simulation the fraction of molecular gas appears to be
converging to approximately 25 \% of the total mass or equivalently 50 \% of the
100 K gas ($2.5 \times 10^8$ M$_{\odot}$). These figures are however dependent
on the scale height used for the column densities of HI and H$_2$, which was
assumed to be 100 pc (the scale height of the disc). Taking a scale height of 25
pc (comparable to the nearest B star from the Sun) reduces the total mass of
H$_2$ by 30\%.  
  
Fig.~5 shows the
density of the 100 K gas (total HI and predicted H$_2$), the predicted H$_2$
density and the $10^4$ K gas density as a function of azimuth. These are the
average densities, calculated over a ring centred at 7.5 kpc, of width 200 pc
and divided into 100 segments.
This figure again shows the much higher density of H$_2$ in the inter-arm regions
compared to the single-phase medium (Fig.~10, \citet{DBP2006}) 
with the density of H$_2$ now 1/10th of
the total cold HI + H$_2$ gas in the inter-arm regions. The mass of molecular gas
is now approximately 1/3 of the total mass of cold gas in 
the spiral arms, and 1/5 in the inter-arm regions. 
Fig.~5 further shows the difference in location of the shocks, as the
density of the hot gas peaks at a lower azimuth than the cold gas.

\begin{figure*}
\centering
\begin{tabular}{c c}
\psfig{file=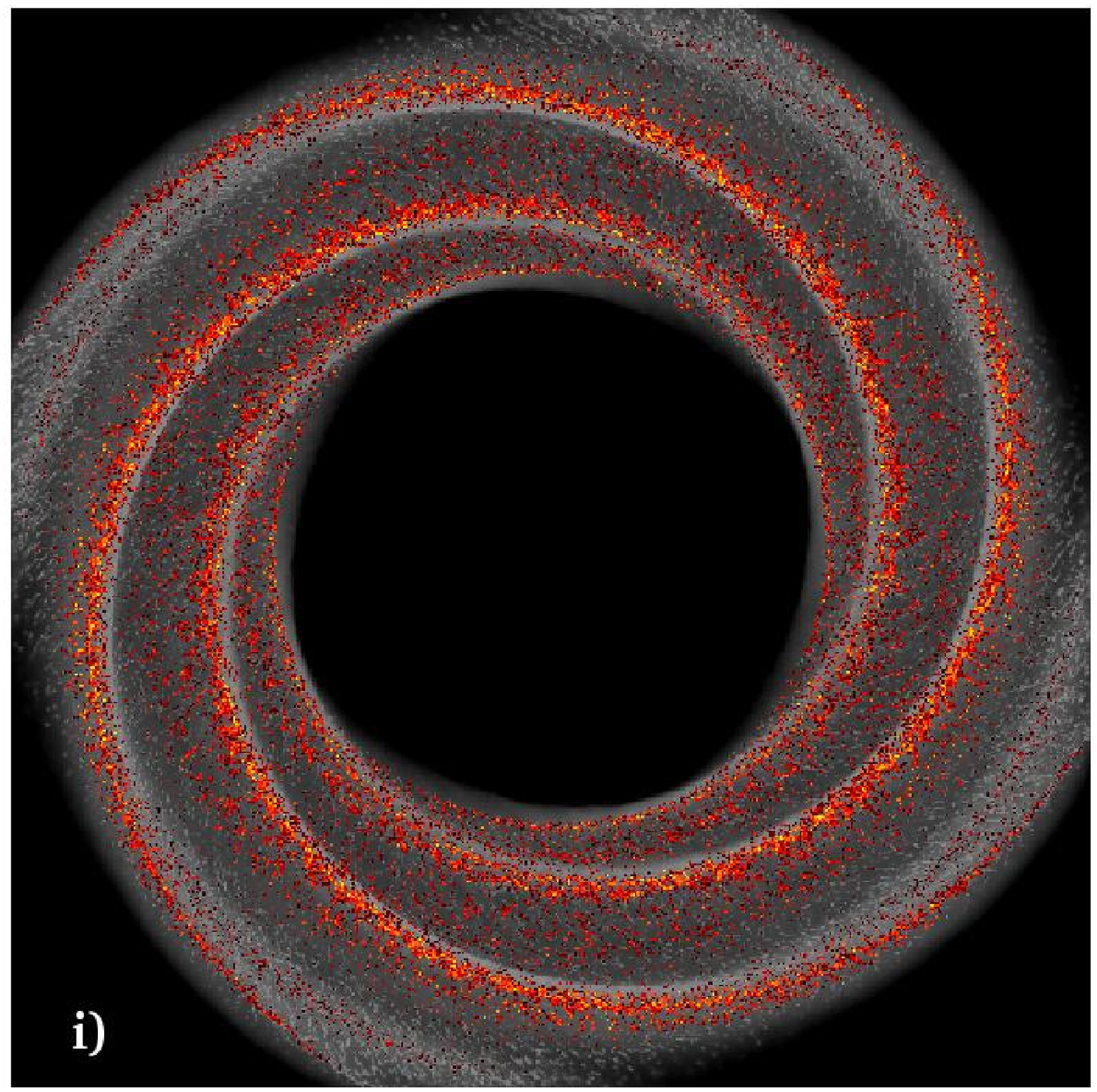,width=2.7in} & 
\psfig{file=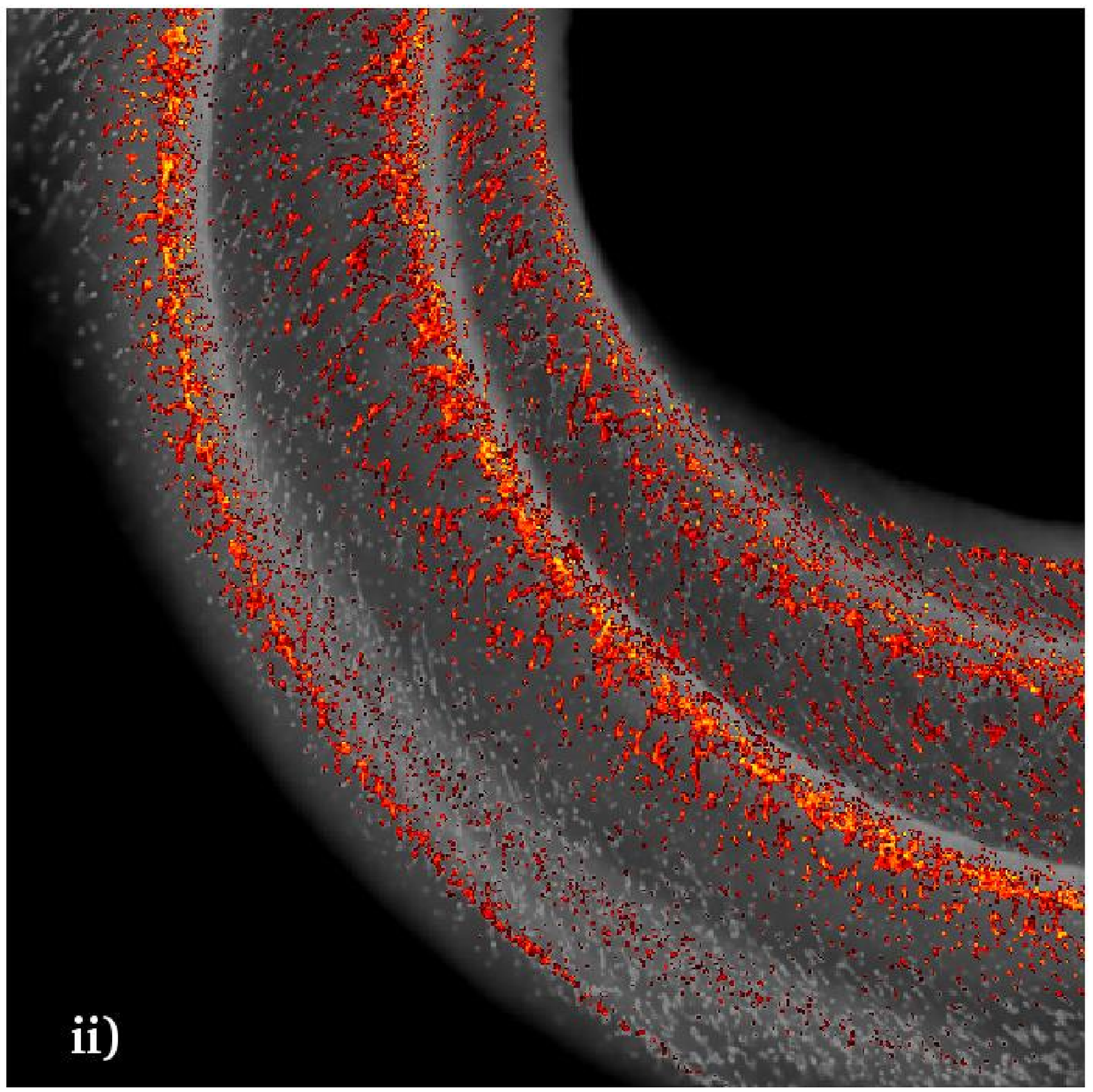,width=2.7in} \\
\psfig{file=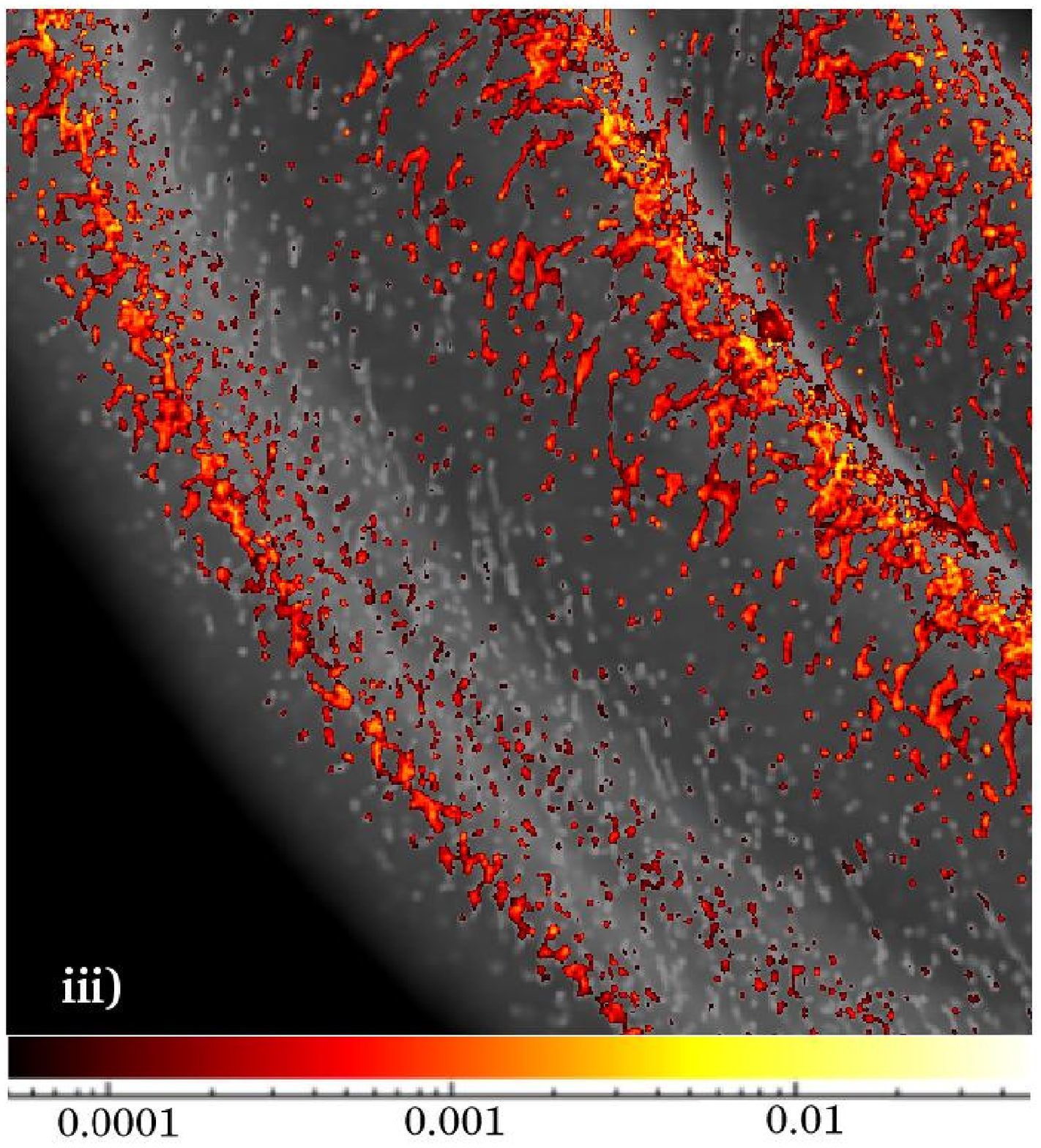,width=2.74in} & 
\psfig{file=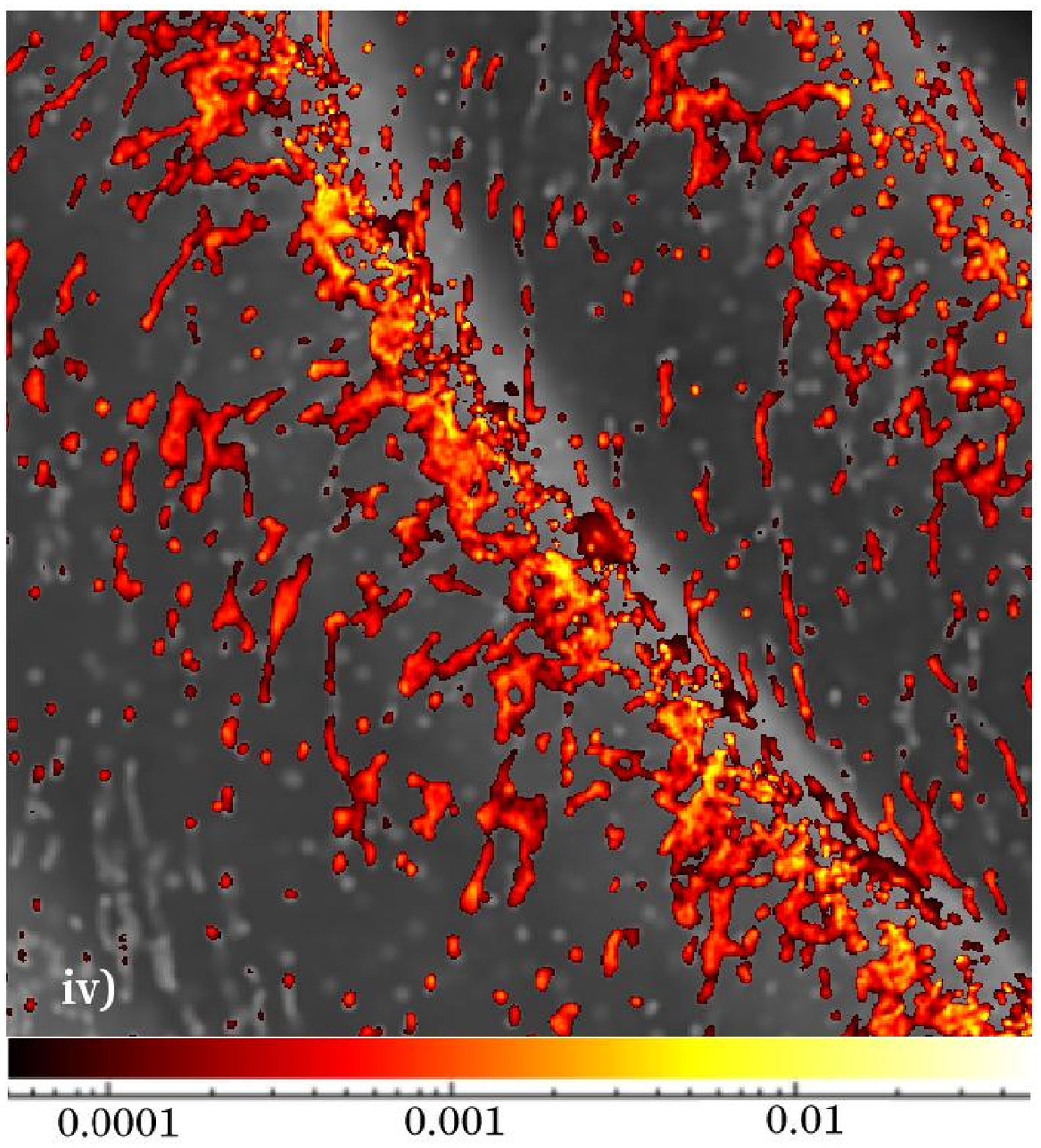,width=2.74in} 
\end{tabular}
\caption{Column density plots (g~cm$^{-2}$) 
showing density of molecular hydrogen (red) against
overall density (black and white).
Length-scales of plots are i) 20 kpc $\times$ 20 kpc, ii) 10 kpc $\times$ 10 kpc,
iii) 5 kpc $\times$ 5 kpc (-8.5 kpc $<x<$ -3.5 kpc, -8.5 kpc $<y<$ -3.5 kpc), 
iv) 3 kpc $\times$ 3 kpc (-6.5 kpc $<x<$ -3.5 kpc, -6.5 kpc $<y<$ -3.5 kpc).
The xy-coordinates assume a Cartesian grid centred on the midpoint of the disc 
which remains fixed with time.
Gas is flowing clockwise across the disc.}
\end{figure*}

\begin{figure}
\centering{
\psfig{file=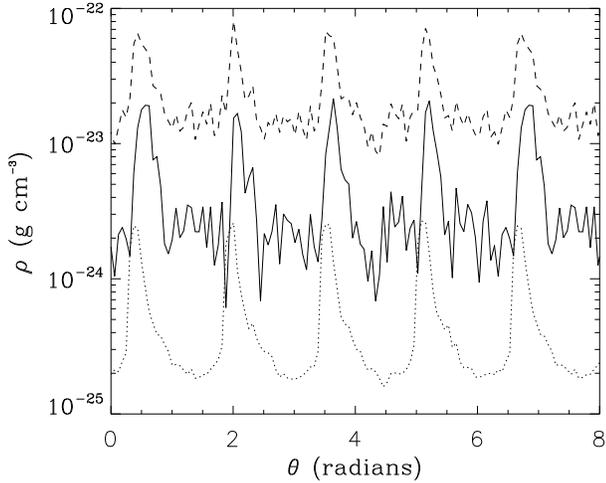,height=2.8in}}
\caption{The average density is plotted (top) versus azimuth, calculated for a 
ring centred at 7.5 kpc ($\theta$ is measured clockwise round the disc). 
The density of the $10^4$ K gas (dotted), 100 K (dashed) 
(including HI and
H$_2$) and H$_2$ (solid) are shown. The bottom figure shows the density of
molecular hydrogen for individual particles (of similar phases) with time.}
\end{figure}

\begin{figure}
\centering{
\psfig{file=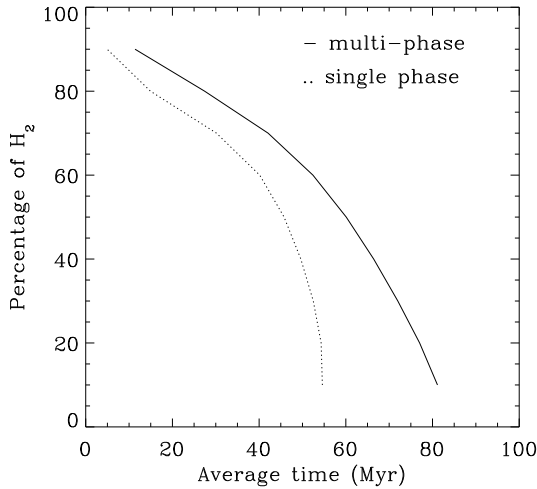,height=2.8in}}
\caption{The typical time over which gas contains over a certain fraction of 
H$_2$ is
shown in this figure. For each percentage of H$_2$, the average time is
calculated from all particles which exceed that fraction of H$_2$ during each
simulation. The average dynamical time for gas exceeding
90\% H$_2$ is 5 Myr, increasing to 20 Myr for gas exceeding 10\% H$_2$.}
\end{figure}

Fig.~6 shows the typical timescale gas spends with different fractions of
H$_2$. The average time for each fraction of H$_2$ is determined from the
duration particles exceed that fraction of H$_2$ during each simulation. The
typical time gas contains over 50\% molecular hydrogen is around 40 Myr for the
single phase simulation and 60 Myr for the multi-phase simulation. These times
decrease to approximately 5 and 10 Myr for gas which contains over 90\% molecular 
hydrogen. When a scale height of 25 pc is used, these timescales decrease by
around 40\%.   

\subsection{Inter-arm atomic and molecular gas}
The distribution of mass with 
density is also shown specifically for the cold component in the inter-arm regions in Fig.~7. 
Clearly the inter-arm gas in the multi-phase simulation reaches densities over a magnitude
higher than for the single phase simulation. 
From local simulations which include cooling, gas with densities $>10$ cm$^{-3}$
is found to be cold, i.e. 100-200 K or less (Fig.~1, \citealt{Glover2006a}). We therefore 
expect  a significant proportion of the cold gas entering the spiral arms to remain cold in 
a multi-phase medium. By contrast, for the single phase calculation, all the gas in the inter-arm 
regions is likely to increase in temperature above 100 K.  
\begin{figure}
\centering
\psfig{file=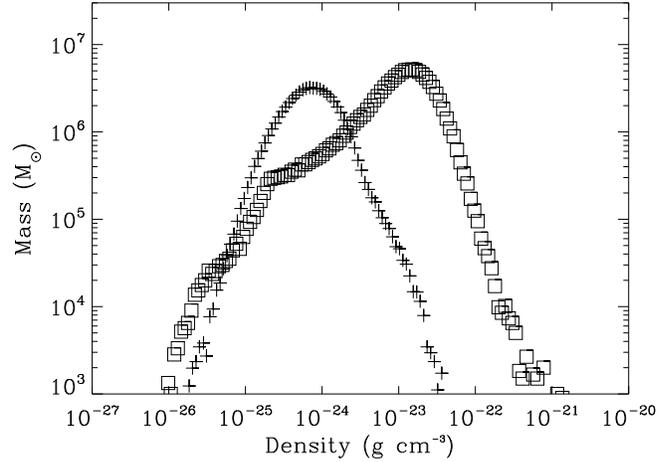,height=2.7in} 
\caption{The distribution of mass as a function of gas density is shown for 
the cold gas in the inter-arm regions. 
The points are determined after
100 Myr for the single (crosses) and multi-phase (squares)  simulations. 
Gas with densities 
above 10 cm$^{-3}$ (10$^{-23}$ g cm$^{-3}$)
 is expected to be cold (100-200 K) \citep{Glover2006a}.}
\end{figure}

For the multi-phase simulation, 
the density of the 100K gas (inclusive of H$_2$) is on average $1-2 \times 10^{-23}$ g 
cm$^{-3}$ and $7 \times 10^{-23}$ g cm$^{-3}$ for
the inter-arm and spiral arm regions respectively (Fig.~5). 
\citet{Allen2004} find from models of photodissociaion regions (PDRs)
that CO(1-0) emission is detected when the density is
10 cm$^{-3}<n<10^5$ cm$^{-3}$. Whilst the spiral arm densities are high enough to
be detected in CO,
much of the inter-arm gas in our simulation lies below this regime 
and would be difficult to detect observationally.
\citet{Kaufman1999} also show that the CO to H$_2$ 
conversion factor  is 
roughly constant for a 100 cm$^{-3}$ 
cloud in a low UV field, providing the column density is 
$\gtrsim 4 \times 10^{21}$ cm$^{-2}$ (comparable with the column density 
threshold used for our clump finding algorithms). 
Below these densities, or with a higher UV field,
H$_2$ is significantly underestimated from CO observations since CO is dissociated more 
readily than H$_2$.

Fig.~4 further indicates that the column density of the inter-arm 
molecular gas is typically between $5\times10^{-4}-10^{-3}$ g cm$^{-2}$.
This is an order of magnitude less than the column densities of H$_2$ observed for HISA clouds which
are associated with CO emission \citep{Klaassen2005}. However 
\citet{Klaassen2005} also examine HISA clouds where there is no CO emission, but infer, 
by means of radiative transfer techniques, column densities of H$_2$ for these clouds of around
$10^{-3}$ g cm$^{-2}$. We postulate that the hot component of the ISM confines clouds 
of cold HI to sufficient densities to maintain a low level of H$_2$. The densities of these clouds 
imply that they will be largely
undetected at present, but are high enough that the clouds are likely to be cold.
We note however that the UV flux in these calculations
is kept constant and we do not include feedback. 
 
\subsection{Resolution effects}
We performed further simulations with 1 million (i.e. 500,000 cold gas particles for the multi-phase case) and 250,000 particles to investigate the effects of numerical resolution.
The lower resolution results in lower peak gas densities since 
with fewer particles the spiral shock is less well resolved.
Compared with Fig.~3., the peak densities for the cold gas decrease by a factor of 3 when 1 million particles are used, and a factor of 8 for 250,000 particles for the multiphase simulations. In the single phase simulations, the peak densities are similar for 1 million particles compared to 4 million, but decrease by a factor of 3 for 250,000 particles.
Consequently the amount of molecular gas decreases with resolution, and only a few per cent of the gas is molecular when there are 250,000 particles.

For the multi-phase simulations, there is also a large decrease in the inter-arm densities of the gas, the peak densities shown on Fig.~7 decreasing by a factor of 10 in the lower resolution simulations. This reflects that these densities are dependent on the densities reached in the spiral arms and the confinement of this gas by the hot phase as it traverses the inter-arm regions. 
The average H$_2$ densities similarly decrease by a factor of 2-3 in the spiral arms compared with 
Fig.~5, and a factor of up to 10 in the inter-arm regions.
There is little change in the the inter-arm densities for the single phase results (Fig.~7) where most gas leaving the spiral arms quickly diffuses to low densities.
Apart from the change in the peak densities, the overall distribution of densities, as shown in Figures~3 and 7, is similar at different resolutions. However the peak densities, as well as molecular gas densities, represent lower limits both in the arm and inter-arm regions.
 
\subsection{Structure in the cold gas}
We now consider quantitatively the structure in the cold gas for the multi-phase 
medium, and compare this with the single-phase cold gas. 
We compare the properties
of clumps from each simulation, using 2 clump finding algorithms. Both
algorithms are grid based, assuming a 2D grid over the galactic disc. The first
(hereafter CF1) selects clumps solely using a density threshold, whilst the 
second (hereafter CF2) uses neighbour lists to assign the
most dense grid cells and their neighbours to clumps. 
For CF1, the density in each grid cell is calculated and all the 
cells above a certain density threshold are selected. 
Cells that are adjacent are
grouped together and particles within these cells classed as a clump. 
CF2 was provided by Paul
Clark and is based on the clump-find method of \citet{Williams1994}. 
The SPH densities are first smoothed onto a 2D grid.
The grid cells are
then ordered by decreasing density. For each cell in turn, the cell 
and its neighbours (subsequently removed from the list) are identified as a 
clump, or assigned to a previously defined clump. Again a density threshold was
set which removed low density cells. The properties of each clump
are then calculated from the grid cells rather than the actual SPH particles.
Since CF2 selects the most dense particles first, density peaks tend to be separated
into separate clumps where they would otherwise constitute 
a single clump for CF1.
In both cases we used the density of the molecular gas, which as seen from
Fig.~4b largely reflects the underlying structure, and so we can compare the
properties of these clumps with observed molecular clouds. 

\begin{figure}
\centering
\begin{tabular}{c}
\psfig{file=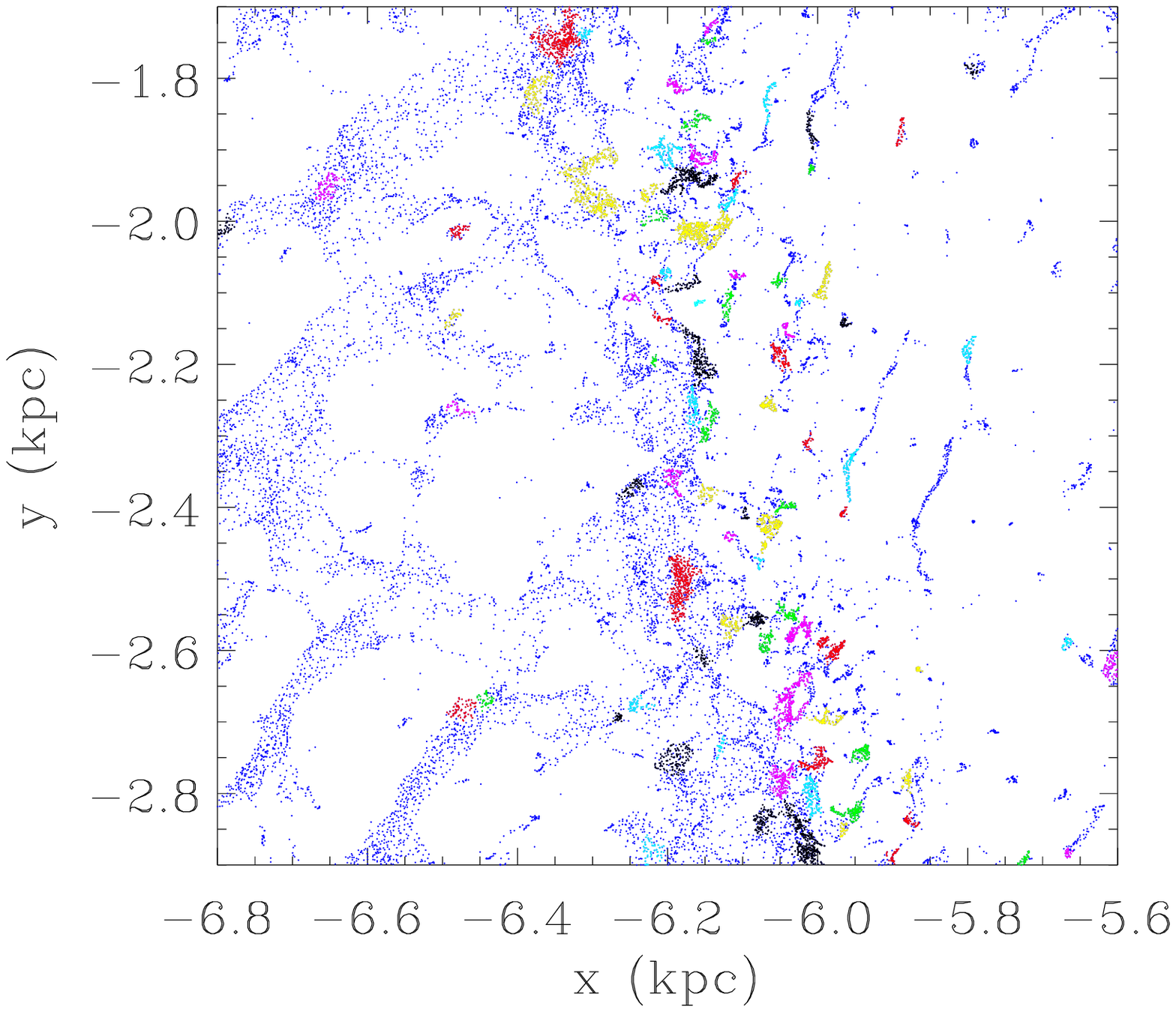,height=3.in} \\
\hspace{0.5in}\psfig{file=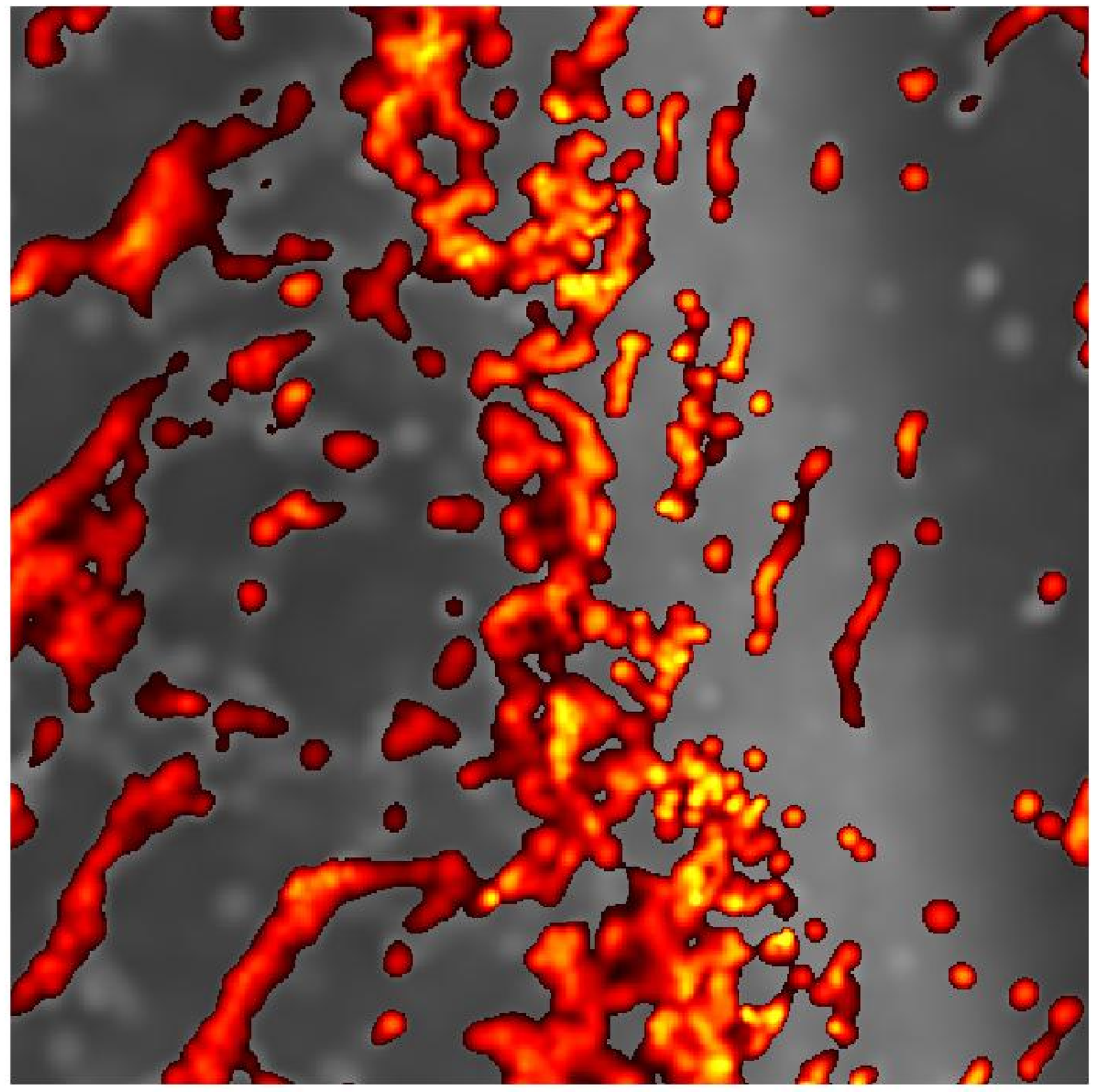,height=2.4in}
\end{tabular}
\caption{The top panel shows the distribution of clumps found using the
clump-finding algorithm CF1 for a 1.2 by 1.2 kpc subsection of the disc after
100 Myr. The particles corresponding to
the 100 K gas are plotted in dark blue, the $10^4$ gas is not shown. The
particles in each clump are over-plotted in a different colour. 
The lower panel shows the corresponding column
density image, using the same scale as Fig.~5 with molecular hydrogen
overplotted in red.}
\end{figure}

\begin{figure} 
\centering
\begin{tabular}{c}
\psfig{file=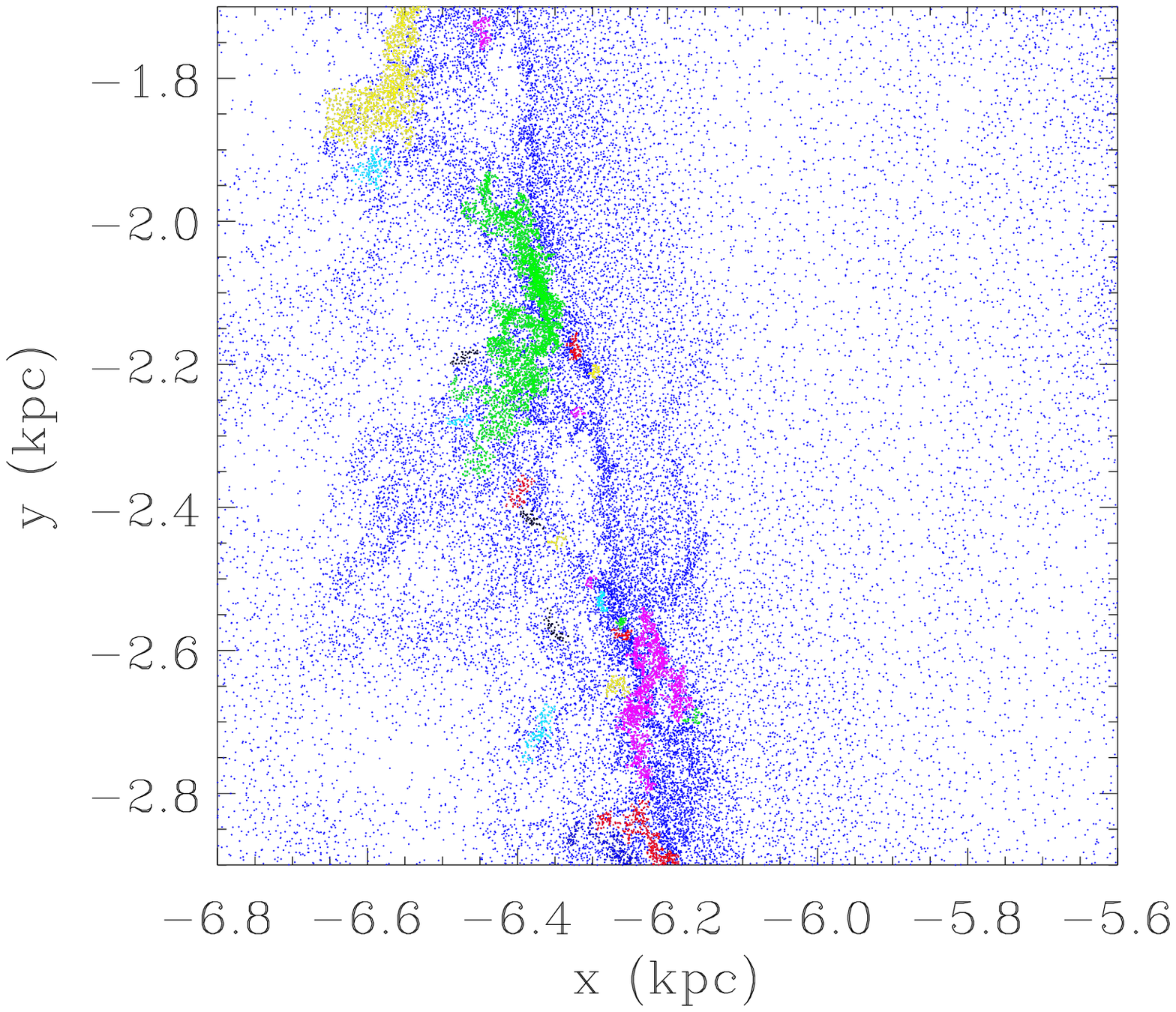,height=3.in} \\
\hspace{0.5in}\psfig{file=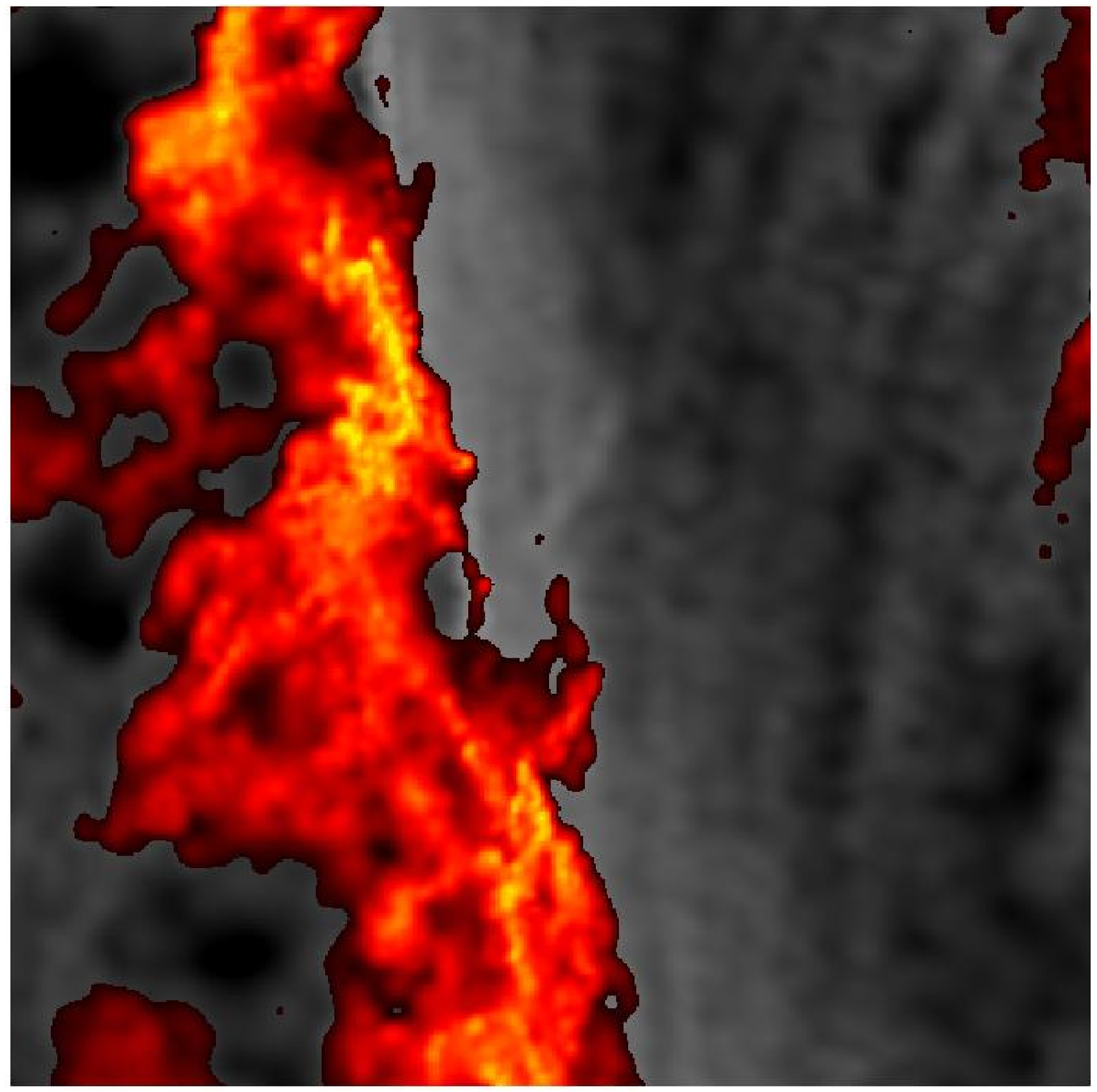,height=2.4in}
\end{tabular}
\caption{The top panel shows the distribution of clumps found using the
clump-finding algorithm CF1 for a 1.2 by 1.2 kpc subsection of the disc after
100 Myr. This is taken from the single-phase
(50 K) simulation. The dark blue points show all the particles in the 
simulation, but the particles in clumps are overplotted in different 
colours. 
The lower panel shows the corresponding column
density image, using the same scale as Fig.~4.}
\end{figure}
The clumps found using CF1 are displayed for the multi-phase (Fig.~8) and
the single-phase medium (Fig.~9) for a subsection of the disc (the same
section is shown in Fig.~2). Both these figures are taken after 100 Myr. 
The resolution of the grid for the algorithm is 5~pc, and the density 
threshold for each cell is 100 M$_{\odot}$, i.e. equivalent
to a surface density of $9 \times 10^{-4}$ g cm$^{-2}$. Each clump consists of at least
30 SPH particles (corresponding to a total hydrogen mass of 5000 M$_{\odot}$)
which lie within cells above the density threshold. Clumps with fewer particles
are discarded.
We also show the corresponding column density plots indicating the molecular
gas column density (Figs~8 and 9, lower).

There is a clear difference in structure of this section of the disc for the 
single and multi-phase gas. The multi-phase case shows many smaller clumps along
the spiral arm, whilst the single-phase plot consists of 3 large clumps and
several smaller clumps. Over the whole disc, there are 8383 clumps in the multi-phase
calculation and 2066 in the single phase calculation.
The clump finding
algorithm also picks out several clumps in the inter-arm regions in the
multi-phase simulation, whereas there are none for the single-phase results. 
These figures
further show the general distribution of cold gas. In the multi-phase 
simulation, the cold gas occupies a much smaller filling
factor, located in small dense clumps, particularly in the inter-arm regions. 
By contrast, in the single-phase case, the cold gas is much more
uniform in the inter-arm regions. The second algorithm, CF2, generally produced 
smaller clumps, but with similar results. 

\subsection{Properties of molecular gas clumps}
We now examine the properties of the clumps from the single and multi-phase 
simulations. The results shown use CF1, but we discuss any differences with the
second clump-finding algorithm. Again only clumps with $> 30 $ particles are retained,
and the clumps are located after 100 Myr.
The mass and radii for the clumps from
the 2 simulations are shown in Fig.~10.  
At later times the clump masses
increase for the multi-phase simulation increase,
as there is a higher fraction of H$_2$, while those of the single phase
simulation are similar.

For both clump-finding algorithms, we take a column density threshold 
$9 \times 10^{-4}$ g cm$^{-2}$.
For CF1, this corresponded to taking a grid of resolution 5
pc and requiring that each cell contains at least 100 M$_{\odot}$ (of H$_2$). 
For CF2, we apply the same grid resolution, and a minimum column density of 
0.001 g cm$^{-2}$ of H$_2$.
With CF1, each clump corresponds to a group of SPH
particles. The radii shown in Fig.~10 are determined by calculating
the radius which contains 3/4 of the total clump mass, and thus the mass plotted
is 3/4 of the total clump mass assigned by the particles. Using the full or 1/2 
clump radii and masses merely shifts the distribution of clumps 
to higher or lower masses and radii.
For CF2, we use the
effective radius from the grid cells which make up each clump, i.e.
$r_{eff}=\sqrt{Area/\pi}$ using the maximum extent of the $x$ and $y$
coordinates to calculate the area.

Fig.~10 shows the mass versus radii from the clumps found in 
the multi and single phase simulations from a quarter of the disc.
The average smoothing length over all the
particles is 30 pc, but reduces to 10 pc for particles above the cutoff density used for the
clump-finding algorithm.   
There are far fewer clumps in the single-phase simulation - less than 1/3 of
the multi-phase simulation. However there are comparatively more larger clouds, as 
also indicated by Figs~8 and 9. 
The total mass of the clumps in Fig.~10 for the single-phase 
simulation is $5.7 \times 10^6$ M$_{\odot}$ compared to $1.4 \times 10^7$ 
M$_{\odot}$ for the multi-phase
simulation, over a quarter of the disc. 
The most massive clump after 100 Myr
in both simulations was $\sim 3 \times 10^5$
M$_{\odot}$, with a radius of 150 pc. 
For the  multi-phase simulation, 
the fraction of molecular hydrogen
increases with time, so after 200 Myr, the maximum cloud mass is 
$\sim 6 \times 10^5$ M$_{\odot}$.

The clumps found using the second algorithm, CF2, follow a similar distribution
to that shown on Fig.~10. However there is less scatter and fewer very
massive clouds. The clumps from both simulations, and from both algorithms 
follow a $M \propto r^{2}$ dependence, implying the clouds have constant surface
densities of $\sim 8$ M$_{\odot}$ pc$^{-2}$ ($2 \times 10^{-3}$ g cm$^{-2}$). 
For comparison, the density
criterion of 4 M$_{\odot}$ pc$^{-2}$ is also shown. The apparent
constant surface density
of the clumps is a consequence of the density threshold applied to find
the clumps. Increasing the threshold 10 times moves the distribution of 
clumps to correspondingly smaller radii. The surface densities of molecular
clouds from observational results (e.g.
\citealt{Heyer1998,Blitz2006}) could similarly
represent a selection effect based on the threshold antenna radiation
temperature, $T_R^*$. 

Since we do not include any feedback effects, most of the clumps are situated in the 
mid-plane of the disk, with average scale heights of $<10$ pc. 
However approximately 10\% of the clouds, generally those of lower mass, have a centre of mass between 50 and 100 pc above or below the plane of the disk.

\begin{figure}
\centering
\begin{tabular}{c}
\psfig{file=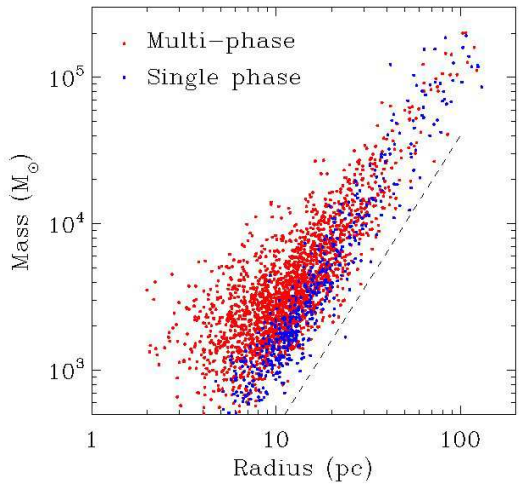,height=2.8in} \\ 
\end{tabular}
\caption{The mass and radius of the clumps are plotted using CF1 over a quarter 
of the disc. Lower mass clumps continue this distribution, but are not shown.
The clumps are selected at 100~Myr. 
The dashed line indicates $\Sigma= 9 \times 10^{-4}$ g cm$^{-2}$ (4 M$_{\odot}$ pc$^{-2}$).}
\end{figure}
The resolution of most of the clumps was insufficient to calculate a velocity
dispersion. However the better resolved clumps (with $>100$ particles) contain
velocity dispersions of a few km s$^{-1}$. We also calculated the virial 
parameter,
\begin{equation}
\alpha_G=\frac{5 \sigma^2 r}{G M}.
\end{equation}
for the clumps. In both simulations, most of the clumps are
unbound. However the more massive clumps tend to be more bound, with $\alpha_G$ 
typically $<10$.

Fig.~11 shows the mass
spectrum from the 2 simulations, for the total number of clouds found over the
whole of the disc (of $>30$ particles) with CF1.
Assuming a mass spectrum $dN/dM \propto M^{-\alpha}$, we find for the
multi-phase simulation, 
$\alpha = 2.35\pm0.15$. The clumps from the single-phase 
simulation produce a much shallower spectrum, of $\alpha=1.6\pm0.3$.
At later times, the spectrum for the single-phase simulation is similar,
although the spectrum for the multi-phase simulation  becomes shallower, 
with $\alpha \sim 2.15$ after 200 Myr, since there are more larger clouds at 
later times.
Using CF2, the larger clumps divide into smaller clumps which
truncates the clump masses at $1.5 \times 10^5$ M$_{\odot}$. Consequently the
value of $\alpha$ increases to  $\alpha \sim 3.3$ and $\alpha \sim 2.6$ for the
multi and single-phase simulations respectively.
The form of the mass spectrum thereby appears to depend on the details of the 
clump-finding algorithm. What can be concluded however is that the multi-phase 
medium produces a steeper mass spectrum and fewer high mass clouds. 
Observations indicate that $\alpha$ lies in the range $\alpha=1.5-1.8$
\citep{Solomon1987,Heyer1998}
for the Galaxy and up to $\alpha=2.5$ for external galaxies 
\citep{Blitz2006}.  

Fig.~12 shows the clump mass spectra for the multi-phase simulation with 6 million, 1 million and 250,000 particles. The clumps are again found using CF1, using the same column density threshold
(hence the clumps lie within the distribution in Fig. 10), but increasing the grid-size for the clump-finding algorithm. All the clumps contain $>$ 30 particles.
The slope becomes shallower as the resolution decreases and with 1 million particles the total amount of H$_2$ is significantly reduced to 10\%. 
At lower resolutions, less massive mass clouds cannot be detected, 
although the upper mass is not significantly reduced compared to the higher resolution simulations. 
For single phase calculations the slope of the spectrum was the same for 4 and 1 million particles, but shallower for 250,000 particles. In summary, we stress that  
the results presented here should be taken as lower limits on the fraction of molecular gas present, both in the spiral arms and in the interarm regions. 
\subsubsection{Inter-arm and spiral arm molecular clouds}
As mentioned previously, the multi-phase simulation produces a much greater 
degree of inter-arm molecular gas and a few of the clumps in Fig.~8 are
located
in the inter-arm regions. We show the distribution of molecular
clouds over the whole of the disc, using CF1, in 
Fig.~13. There are
a significant number of clouds in the inter-arm regions, especially in the inner
regions of the disc where the spiral arms are closer together. For the outer
parts of the disc, the inter-arm clouds tend to be situated on the edge of 
spiral arms. We estimate the
ratio of inter-arm to arm clouds is approximately 1:7 (decreasing to 1:10 for the
1 million particle simulation, and no inter-arm clouds at lower resolution). 
The distribution of clouds and the inter-arm ratio is very similar for the
second clump-finding algorithm.
By contrast, for the
single phase simulation, the clouds all lie along the spiral arms. 
Consequently observational-style plots of the molecular clouds over a range of longitudes 
(\citealt{DBP2006} Fig. 14 \& 15) show a greater degree of scatter in the inter-arm regions.

We also considered the number of more massive clouds 
in the inter-arm and arm regions. Approximately 3\% of inter-arm clouds were 
$>10^4$ M$_{\odot}$  compared to 6\% of clouds in the spiral arms.
The cloud mass spectra for clouds located in the spiral arms and inter-arm 
regions are displayed in Fig.~14. 
The spectrum for the spiral arm clouds is very similar to the
spectrum for the total number of clouds, although slightly shallower with
$\alpha = 2.25\pm0.15$. The spectrum for the inter-arm clouds however is much
steeper with $\alpha =2.9\pm0.4$, and a cut off in mass at around 
$3 \times 10^4$ M$_{\odot}$.
\begin{figure}
\centering
\psfig{file=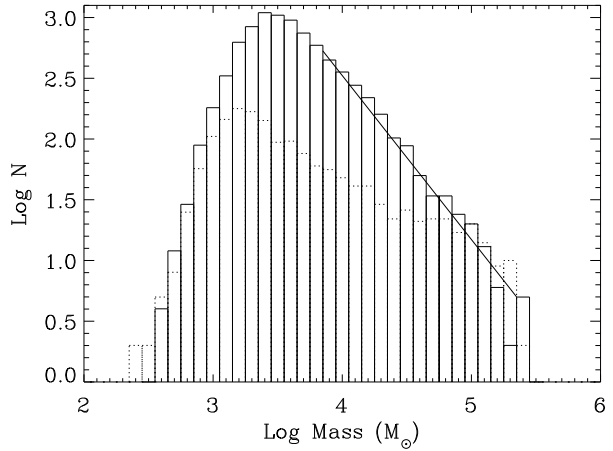,height=2.6in} 
\caption{The clump mass spectrum is shown for clumps 
from the multi-phase
simulation (solid) and the 50 K simulation (dotted). Both are determined 
after 100 Myr and use CF1. The masses indicated are the mass of H$_2$.
The imposed limit of 30 particles corresponds to a total (HI and H$_2$) mass of
approximately $10^{3.5}$ M$_{\odot}$. 
Allowing that the actual mass of H$_2$ is a fraction 
of the total, the resolution limit corresponds 
with the peak of the distributions, below which the sample is 
incomplete and the number of clumps rapidly drops off.  
The slope corresponds to $dN/dM \propto M^{-2.35}$ for clumps from the multi-phase
simulation compared to $dN/dM \propto M^{-1.6}$ for the single phase results.}
\end{figure}

The number of
inter-arm clouds is sensitive to the photodissociation rate, so these figures 
only provide a
rough indication. The photodissociation rate may be lower in the inter-arm
regions, increasing the number of inter-arm clouds. On the other hand, the
photodissociation rate will be higher where massive star formation occurs, which
with the effects of feedback may disrupt molecular gas clouds before
they enter the inter-arm regions.
\begin{figure}
\centering
\psfig{file=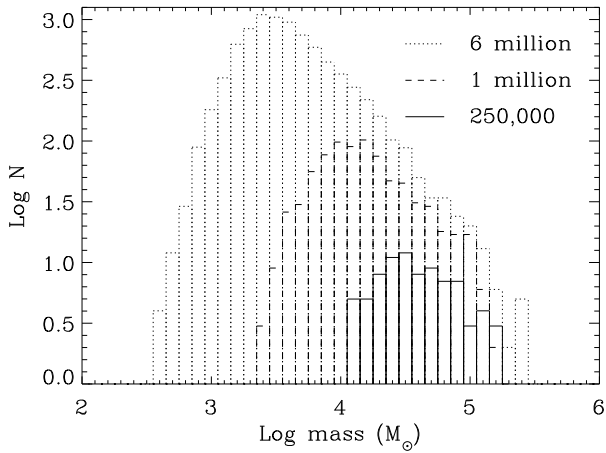,height=2.6in} 
\caption{The clump mass spectrum is shown for the multi-phase simulation using different resolutions.
The clump mass is the mass of H$_2$.
The total number of particles is indicated for each simulation.}
\end{figure}

\begin{figure}
\centering
\begin{tabular}{c}
\psfig{file=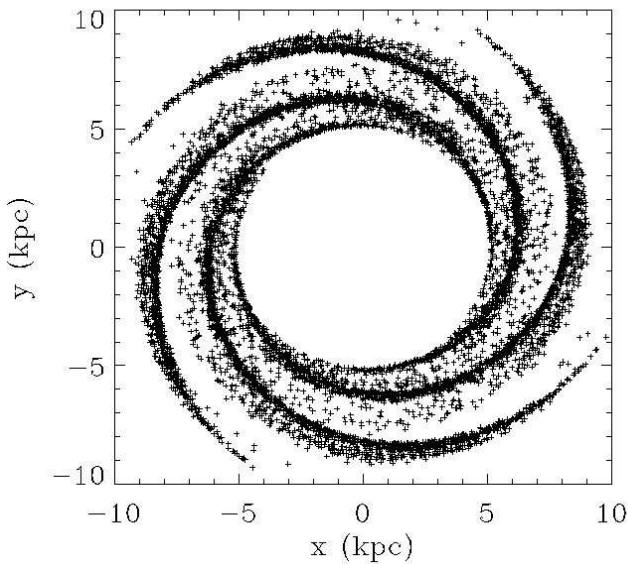,height=3.2in} 
\end{tabular}
\caption{
The distribution of clumps is displayed 
for the multi-phase simulation, using CF1 after 100 Myr.}
\end{figure}

\begin{figure}
\centering
\begin{tabular}{c}
\psfig{file=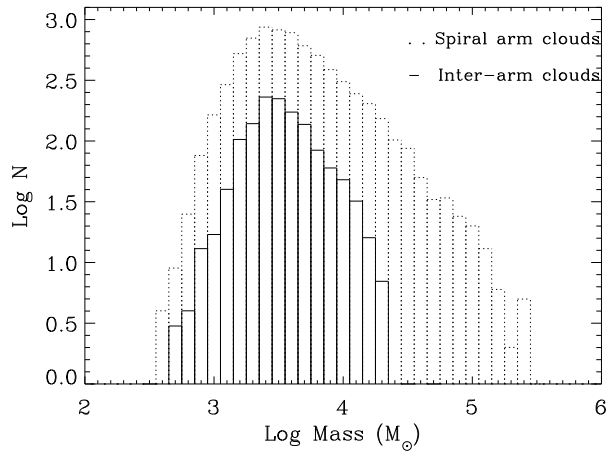,height=2.6in} 
\end{tabular}
\caption{
The clump mass spectrum is shown for the clumps in the spiral arms (dotted) 
and inter-arm regions (solid) from the multi-phase simulation. 
The clumps are selected at 100 Myr, and again
the minimum (total) mass resolution of the clumps is $10^{3.5}$~M$_{\odot}$.}
\end{figure}

\section{Conclusions}
We have performed hydrodynamic simulations of a multi-phase medium subject to a
spiral potential. We find that including hot gas in addition to cold leads to higher densities
in the cold gas. Consequently the amount of molecular gas increases in the disc, with  
1/4 of the cold gas molecular after 100 Myr, approximately double that of
corresponding single phase calculations. This fraction further increases
with time. 
The increase in H$_2$ is most
striking in the inter-arm regions, with molecular clouds surviving into the
inter-arm regions without being destroyed by photodissociation, although 
both feedback, and heating and cooling of the gas are neglected.
For the whole disc, 
approximately 1/7 
of the molecular clouds lie in the inter-arm regions, but this ratio decreases
with radius as the separation of the spiral arms increases.
The spiral arm H$_2$ densities increase to values comparable with average
GMC densities, whilst the typical inter-arm densities are below those detected
with CO. 
We therefore propose that a multi-phase
medium predicts a population of cold clouds of HI and molecular hydrogen,
which are currently undetected. 

The multi-phase simulations also show more structure in the cold gas, with
smaller clumps present compared to the single phase calculations. Consequently
the mass spectrum is steeper for the multi-phase clouds, although the
value of the exponent is found to depend on the clump-finding method.
The maximum cloud mass is $6\times10^5$ M$_{\odot}$ in the multi-phase
simulation, comparable to GMCs in M33 \citep{Engargiola2003}, but an order of
magnitude lower than
the most massive GMCs observed in the Milky Way.
The free-fall times for the molecular clouds are esimated to range from 5 Myr, when
over 90\% of the gas is molecular, to 20 Myr when over 10\% of the gas is molecular. 
The lifetimes of the clouds are estimated to be 20 Myr for 
when over 90\% of the gas is molecular, and 80 Myr when over 10\% of the gas is molecular, both of
which correspond to roughly 4 free-fall times (5 Myr for $>$ 90\% and 20 Myr for $>$ 10\% H$_2$).

The main factor which will affect these results is the inclusion of feedback,
particularly in determining how much molecular gas is retained in the ISM 
following star formation. This may lead to less inter-arm molecular gas, but
would also introduce the possibility of triggered molecular cloud formation in
the inter-arm regions. Furthermore, using lower scale heights to 
determine the photodissociation rate could also reduce the amount of H$_2$..
Thirdly we do not include heating and cooling of the
different phases. Rather we have assumed a fixed composition of the ISM to
approximately agree with observations. Previous simulations of the
ISM with heating and cooling show that a cold phase of gas is sustained, with
20-30\% of gas \citep{Audit2005} to 60\% or more 
\citep{Wada2001,Piontek2005} in the cold 
regime. We leave a more complete treatment of the thermodynamics of the
ISM and spiral shocks for future work.
    
\section*{Acknowledgments}
Computations included in this paper were performed using the UK Astrophysical
Fluids Facility (UKAFF). We would like to thank the referee, Simon Glover, for helpful
suggestions which lead us to look further into the results of our simulations.   

\bibliographystyle{mn2e}
\bibliography{Dobbs}

\bsp

\label{lastpage}

\end{document}